\documentclass[a4paper,11pt]{article}
\usepackage{pos}

\usepackage{graphicx}
\usepackage{dcolumn}
\usepackage{bm}
\usepackage{bbold}

\usepackage{color}
\usepackage{amsfonts}
\usepackage{subfigure}
\usepackage{array}

\newcommand{\Tr}{\ensuremath{\operatorname{Tr}}}

\newcolumntype{L}{>{\centering\arraybackslash}m{3cm}}

\definecolor{bjcol}{rgb}{1,.44,0.13}

\definecolor{blue}{rgb}{0,0,1}

\definecolor{green}{rgb}{0,1,0}

\definecolor{red}{rgb}{1,0,0}

\definecolor{gray}{rgb}{.5,.5,.5}

\definecolor{darkgreen}{rgb}{.0,.5,.0}

\def\Fig#1{Fig.~\ref{#1}} 

\def\Eq#1{Eq.~(\ref{#1})}

\def\eqref#1{(\ref{#1})}

\def\sec#1{Sec.~\ref{#1}}

\def\lA0{{\langle A_0 \rangle}}
\def\bA0{{\bar{A}_0}}

\def\0#1#2{\frac{#1}{#2}}

\graphicspath{{./figures/}{./}}

\title{High-order baryon number fluctuations within the fRG approach}

\author*[a]{Wei-jie Fu}
\author[b]{Xiaofeng Luo}
\author[c,d]{Jan M. Pawlowski}
\author[e]{Fabian Rennecke}
\author[a]{Rui Wen}
\author[a]{Shi Yin}

\affiliation[a]{School of Physics, Dalian University of Technology, Dalian, 116024, P.R. China}
\affiliation[b]{Key Laboratory of Quark \& Lepton Physics (MOE) and Institute of Particle Physics,
		Central China Normal University, Wuhan 430079, China}		
\affiliation[c]{Institut f\"ur Theoretische Physik, Universit\"at Heidelberg, Philosophenweg 16, 69120 Heidelberg, Germany}
\affiliation[d]{ExtreMe Matter Institute EMMI, GSI, Planckstra{\ss}e 1, D-64291 Darmstadt, Germany}
\affiliation[e]{Brookhaven National Laboratory, Upton, NY 11973, USA}

\abstract{We compute high-order baryon number fluctuations at finite temperature and density within a QCD-assisted low energy effective field theory. Quantum, thermal and density fluctuations are incorporated with the functional renormalization group approach. Quantum and in-medium fluctuations are encoded via the evolution of renormalization group flow equations. The resulting fourth- and sixth-order baryon number fluctuations meet the lattice benchmark results at vanishing density. They are consistent with experimental measurements, and in particular, the non-monotonic dependence of the kurtosis of net-baryon number distributions on the collision energy is observed in our calculations. This non-monotonicity arises from the increasingly sharpened chiral crossover with the decrease of collision energy.}

\FullConference{%
	The International conference on Critical Point and Onset of Deconfinement - CPOD2021\\
  	15 – 19 March 2021\\
 	Online - zoom
 }

\begin{document}
\maketitle

\section{Introduction}
\label{sec:int}
	
Recent years have seen significant progress in the studies of QCD phase structure at finite temperature and densities from both the theoretical and the experimental side. The pseudo-critical temperature of the chiral crossover at vanishing baryon chemical potential $\mu_B=0$ with physical quark masses has been determined with high accuracy \cite{Bazavov:2018mes, Borsanyi:2020fev}, and in the chiral limit, the critical temperature has also been extracted \cite{Ding:2019prx, Braun:2020ada}. Lattice simulations of QCD phase boundary, QCD thermodynamics, etc., have been extended to the regime of $\mu_B/T\lesssim 2\sim 3 $ \cite{Bazavov:2018mes, Borsanyi:2018grb, Borsanyi:2020fev, Bazavov:2020bjn, Bollweg:2021vqf, Borsanyi:2021hbk}. Remarkably, first-principle calculations of functional QCD, e.g., functional renormalization group (fRG) and Dyson-Schwinger equations (DSE), by now show converging   predictions for the chiral phase structure of QCD, see \cite{Fischer:2018sdj,Fu:2019hdw,Gao:2020fbl,Gunkel:2021oya}: at $\mu_B=0$ lattice benchmarks are met. Within the regime of quantitative reliability of the current functional computations, $\mu_B/T\lesssim 4$, no critical end point (CEP) is found. In turn, the predictions for the location of the CEP or the onset of new physics converge within the regime $(135\,,\, 450\, )\,\textrm{MeV}\lesssim (  {T}_{_{\tiny{\text{CEP}}}}\,,\,{\mu_B}_{_{\tiny{\text{CEP}}}})\lesssim  ( 100\,,\, 650)\,\textrm{MeV}$, outside the regime of quantitative reliability of the current functional computations.  

In experiments signs for a non-monotonic dependence of the kurtosis of net-proton number distributions on the collision energy are observed with $3.1\,\sigma$ significance for central collisions by STAR collaboration \cite{Adam:2020unf}. The measurements have recently been extended to the sixth-order cumulants \cite{STAR:2021rls}. In this proceeding, we present results from a QCD-assisted low energy effective field theory (LEFT) within the fRG approach to study high-order baryon number fluctuations at finite temperature and baryon chemical potential. This LEFT is benchmarked with lattice results at vanishing chemical potential, and the results at finite density are confronted with experimental data for the fourth- and sixth-order cumulants of net-proton number distributions. Emphasis will be put on the underlying reason accounting for the non-monotonic collision-energy behavior of the high-order baryon or proton number fluctuations. Due to the restriction on the length of this proceeding, more details about this work can be found in \cite{Fu:2021oaw}.

\section{QCD-assisted LEFT within the fRG approach}
\label{sec:fRG-LEFT}

We start with a brief overview on the QCD-assisted low-energy effective field theory used in this work, for more details see \cite{Fu:2021oaw}. Its effective action reads for $N_f=2$ flavors, 
\begin{align}
\Gamma_k=&\int_x \bigg\{Z_{q,k}\bar{q} \Big [\gamma_\mu \partial_\mu -\gamma_0(\mu+i g A_0) \Big ]q+\frac{1}{2}Z_{\phi,k}(\partial_\mu \phi)^2 +h_k\,\bar{q}\,
\left(T^0\sigma+\bm T\cdot\bm{\pi}\right)\,q +V_k(\rho,A_0)-c\sigma \bigg\}\,,\label{eq:action}
\end{align}
where $k$ denotes the renormalization group (RG) scale. Moreover, $\int_{x}=\int_0^{1/T}d x_0 \int d^3 x$, where $T$ is the temperature. Here $q=(u\,,d)^{T}$ and $\phi=\left(\sigma,\bm{\pi}\right)$ are the quark and meson fields, respectively, and $Z_{q,k}$, $Z_{\phi,k}$ are their wave function renormalizations. The quark chemical potential in the flavor matrix is given by $\mu = \mathrm{diag}(\mu_u,\mu_d)$, and $\mu_u=\mu_d=\mu_B/3$ is assumed in this work, with the baryon chemical potential $\mu_B$. In \Eq{eq:action}, $(T^0, \bm T)$ are the $U(2)$ generators in the flavor space with $\Tr(T^{a}T^{b})=\frac{1}{2}\delta^{ab}$ ($a,b\in 0, 1,...3$), and the Yukawa coupling is denoted by $h_k$. The effective potential $V_k(\rho,A_0)$ is invariant under the transformation of $O(4)$ with $\rho=\phi^2/2$, and $A_0$ is the temporal component of the gluon background field. The $c$-term in \Eq{eq:action} breaks chiral symmetry explicitly and its strength determines the current quark mass. The full quantum effective action is recovered from the flow equation at $k \rightarrow 0$.

The first-principle computations for functional QCD \cite{Fu:2019hdw} entail, that the dynamics of 2+1-flavor QCD can be well captured based on that of 2-flavor via a scale matching for the temperature and baryon chemical potential, to wit,
\begin{align}
T^{(N_f=2)} &=c_{_{T}}\,T^{(N_f=2+1)} \,,\qquad
\mu_B^{(N_f=2)} = c_{\mu_B}\,\mu_{B}^{(N_f=2+1)} \,.
\label{eq:rescale}
\end{align}
With the temperature dependence of the fourth-order baryon number fluctuations and the curvature of the chiral phase boundary calculated in both 2+1- and 2-flavor QCD, one is able to determine the coefficients in \Eq{eq:rescale}, as $c_{_{T}}=1.247(12)$ and $c_{\mu_B}=1.110(66)$ \cite{Fu:2021oaw}. The errors here comprise the systematic errors in our theoretical calculations, which are denoted by error bands in our numerical results in \sec{sec:num}.

The thermodynamic potential density for a grand canonical ensemble with temperature $T$ and baryon chemical potential $\mu_B$ reads
\begin{align}
\Omega[T,\mu_B]=&V_{k=0}(\rho,A_0)-c\sigma\,,\label{eq:Omega}
\end{align}
where the sigma meson field $\sigma$ and the temporal gluon field $A_0$, or its closely related counterpart, the Polyakov loop $L(A_0)$, are on their respective equations of motion. With the thermodynamic potential density in \Eq{eq:Omega}, or the pressure with $p=-\Omega$, one is able to obtain baryon number fluctuations of a given order $n$, as follows
\begin{align}
\chi_n^{B}&=\frac{\partial^n}{\partial (\mu_B/T)^n}\frac{p}{T^4}\,.\label{eq:suscept}
\end{align}
In experiments it is more advantageous to make use of the ratio between the $n$- and $m$-th order fluctuations, i.e.,
\begin{align}
R_{nm}^{B}&=\frac{\chi_n^{B}}{\chi_m^{B}}\,.
\label{eq:Rnm}
\end{align}
%

\section{Numerical results and discussions}
\label{sec:num}

%
\begin{figure*}[t]
\centering
\includegraphics[width=0.8\textwidth]{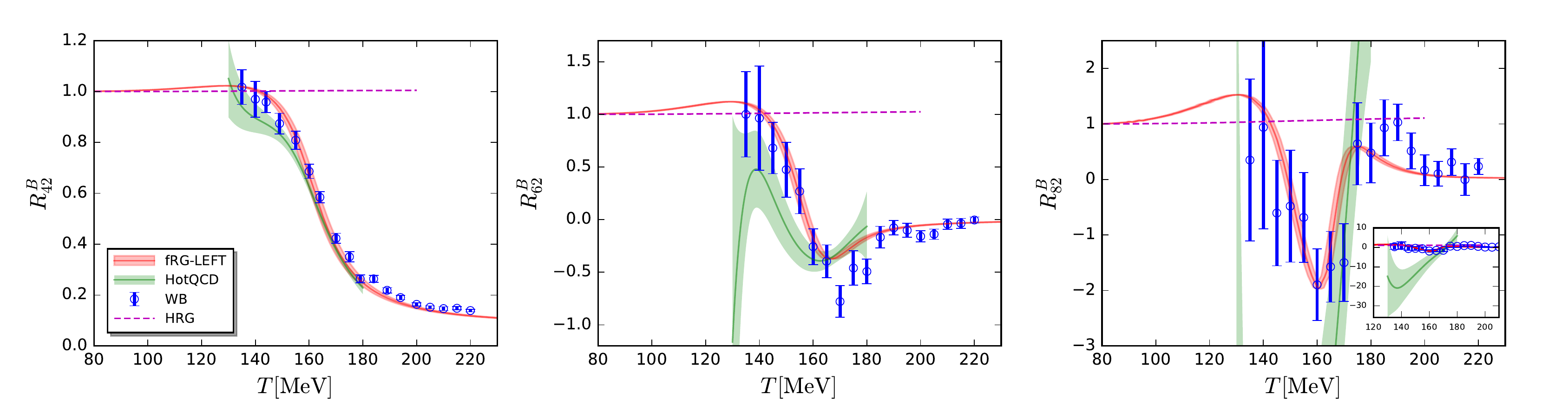}
\caption{Dependence of $R^{B}_{42}=\chi^{B}_{4}/\chi^{B}_{2}$ (left), $R^{B}_{62}=\chi^{B}_{6}/\chi^{B}_{2}$ (middle), and $R^{B}_{82}=\chi^{B}_{8}/\chi^{B}_{2}$ (right) on the temperature with $\mu_B=0$. We compare the results of fRG-LEFT with those of lattice QCD by the HotQCD collaboration \cite{Bazavov:2020bjn} and the Wuppertal-Budapest collaboration (WB) \cite{Borsanyi:2018grb}. The inset in the plot of $R^{B}_{82}$ shows its zoomed-out view. Results of a hadron resonance gas (HRG) model \cite{BraunMunzinger:2003zd} are also presented for comparison.
}\label{fig:R42R62R82-T-muB0}
\end{figure*}
%

%
%

%
\begin{figure*}[t]
\centering
\includegraphics[width=0.44\textwidth]{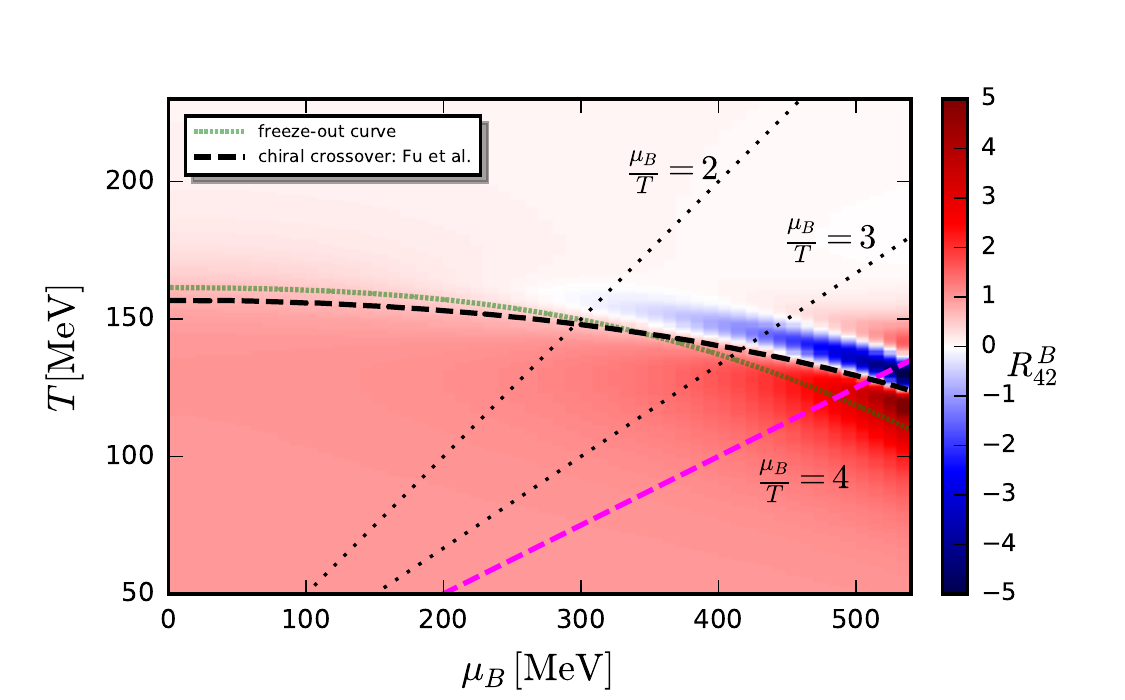}
\includegraphics[width=0.35\textwidth]{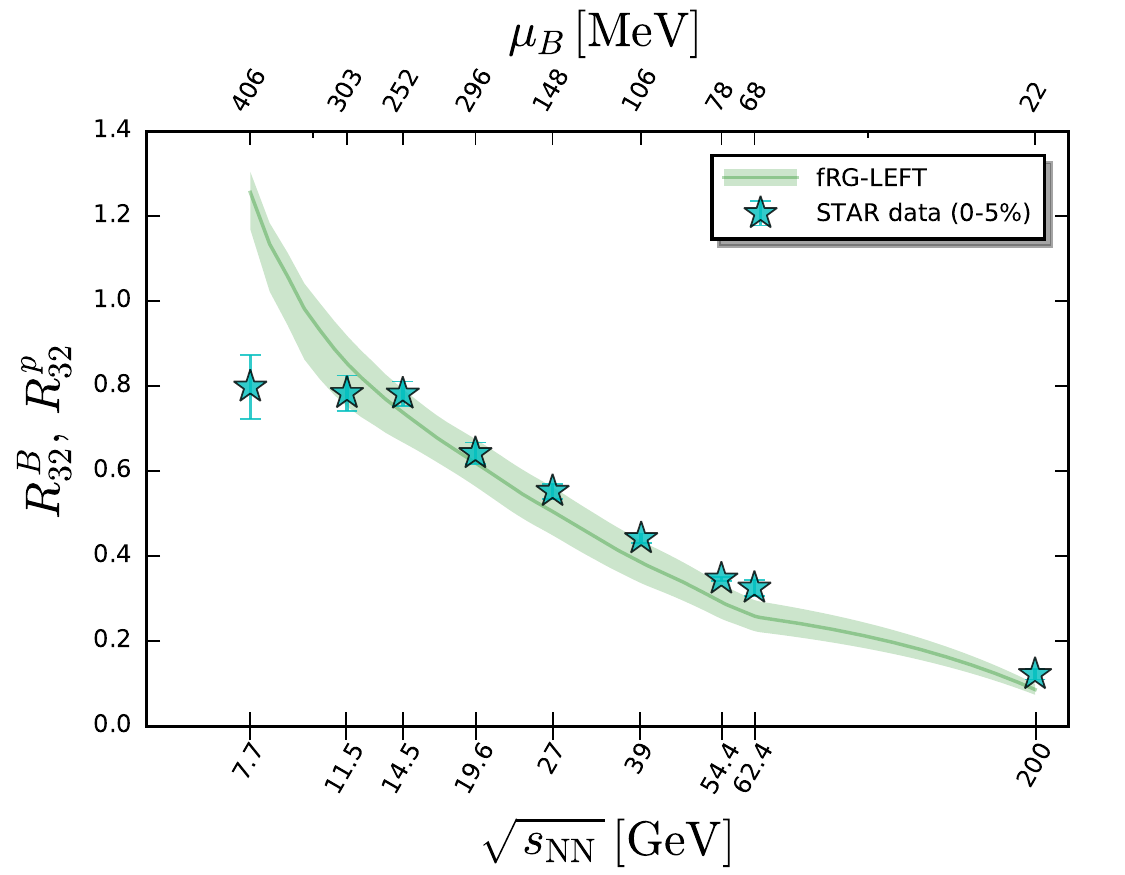}
\caption{Left: Quartic baryon number fluctuation $R^{B}_{42}(T,\mu_B)$ calculated within fRG-LEFT in the phase diagram spanned by the temperature and baryon chemical potential. The black dashed line denotes the chiral crossover of $N_f=2+1$-flavor QCD obtained in a first-principle fRG-QCD computation \cite{Fu:2019hdw}. The green dotted line represents the freeze-out curve, and lines $\mu_B/T=2,3,4$ are also depicted. 
Right: Ratio of cubic to quadratic baryon number fluctuations $R^{B}_{32}$ as a function of the collision energy obtained with the freeze-out curve shown in the left panel. Experimental data for the skewness of the net-proton distributions $R^{p}_{32}$ with centrality 0-5\% \cite{Adam:2020unf} are presented for comparison.}\label{fig:phasediagram}
\end{figure*}
%

In \Fig{fig:R42R62R82-T-muB0} we show the fourth-, sixth-, and eighth-order net-baryon number fluctuations divided by the quadratic one as functions of the temperature at zero baryon chemical potential. Our calculated results in the QCD-assisted LEFT within the fRG approach are compared with the lattice results by the HotQCD collaboration \cite{Bazavov:2020bjn} and the Wuppertal-Budapest collaboration (WB) \cite{Borsanyi:2018grb}. Note that the lattice results of $R^{B}_{62}$ and $R^{B}_{82}$ have not yet been extrapolated to the continuum limit, and there is still significant discrepancy for them between the two lattice collaborations. Our results are in quantitative accordance with those from WB, and qualitatively consistent with those from HotQCD. In the low temperature regime, the net-baryon number obeys the Skellam distributions that are well described by the hadron resonance gas (HRG) model. 


In the left panel of \Fig{fig:phasediagram} the fourth-order baryon number fluctuation $R^{B}_{42}$ is depicted in the phase diagram, whose values are denoted by gradient color. The fluctuations are computed in the QCD-assisted LEFT with fRG. One observes that there is a narrow blue band in the crossover regime, which corresponds to a valley structure of $R^{B}_{42}$ there. 
In this phase diagram we also show the chiral crossover line obtained in a first-principle fRG-QCD computation \cite{Fu:2019hdw}, and the blue band just sits upon the crossover line. It is obvious that the color standing for value of $R^{B}_{42}$ deepens with the increase of $\mu_B$, that is, the chiral crossover is sharpened. In the phase diagram we also show the freeze-out curve, obtained from the freeze-out parameters in STAR experiment \cite{Adamczyk:2017iwn}, in combination with some general considerations, cf. \cite{Fu:2021oaw} for more details. The freeze-out curve in this proceeding is one of freeze-out curves investigated in \cite{Fu:2021oaw}, denoted originally as STAR Fit II, which serves as the best-informed one. In the right panel of \Fig{fig:phasediagram} we show skewness of the baryon number distribution as a function of the collision energy obtained in the fRG-LEFT, confronted with the experimentally measured skewness of the net-proton distributions $R^{p}_{32}$ with centrality 0-5\% \cite{Adam:2020unf}. The freeze-out curve in the left panel of \Fig{fig:phasediagram} has been employed to extract the collision-energy dependence of $R^{B}_{32}$. One finds that the theoretically calculated results of $R^{B}_{32}$ are in good agreement with the experimental data of $R^{p}_{32}$ with $\sqrt{s_{\rm NN}}\gtrsim 14.5$ GeV. However, there is a significant discrepancy in the regime of low collision energy, which might indicate that the effect of global baryon number conservation begins to play a role there. 

%
\begin{figure*}[t]
\centering
\includegraphics[width=0.6\textwidth]{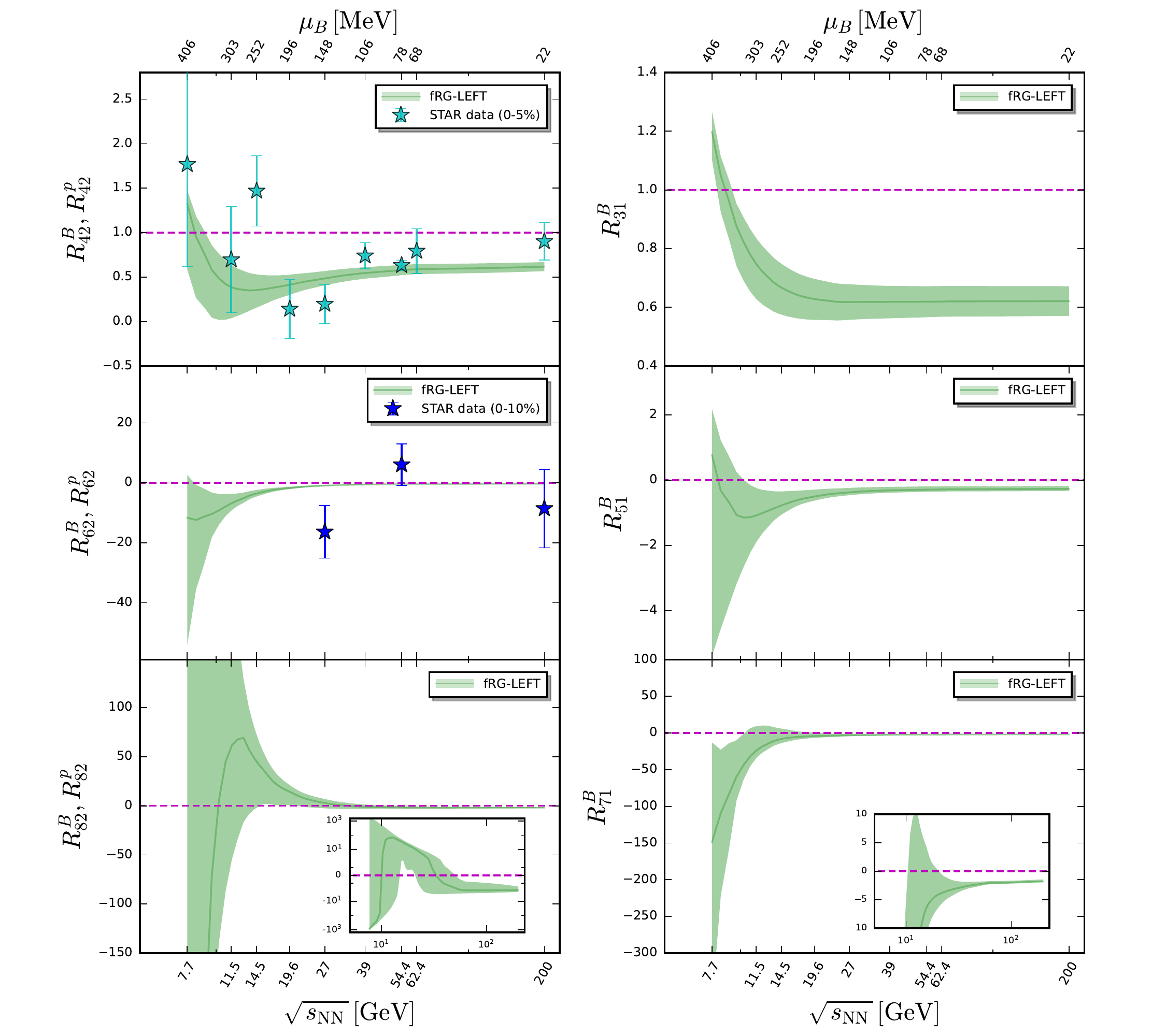}
\caption{Baryon number fluctuations of even orders $R^{B}_{42}$ (top-left), $R^{B}_{62}$ (middle-left), $R^{B}_{82}$ (bottom-left) and those of odd orders $R^{B}_{31}$ (top-right), $R^{B}_{51}$ (middle-right), $R^{B}_{71}$ (bottom-right) as functions of the collision energy obtained in the fRG-LEFT. Theoretical results of $R^{B}_{42}$ are compared with the kurtosis of the net-proton number distributions $R^{p}_{42}$ measured in heavy-ion collisions with centrality 0-5\% \cite{Adam:2020unf}, and $R^{B}_{62}$ with the sixth-order cumulant of the net-proton number distributions $R^{p}_{62}$, which is measured at three values of collision energy, i.e., $\sqrt{s_{\mathrm{NN}}}$=200 GeV, 54.4 GeV and 27 GeV with centrality 0-10\% \cite{STAR:2021rls}. 
}\label{fig:Rm2Rm1-sqrtS}\vspace{-0.5cm}
\end{figure*}
%

In \Fig{fig:Rm2Rm1-sqrtS} we show the dependence of baryon number fluctuations $R^{B}_{42}$, $R^{B}_{62}$, $R^{B}_{82}$, $R^{B}_{31}$, $R^{B}_{51}$, $R^{B}_{71}$ on the collision energy obtained in the QCD-assisted LEFT within the fRG approach, where the freeze-out curve in the left panel of \Fig{fig:phasediagram} has been employed. In the same time, in the plots of $R^{B}_{42}$ and $R^{B}_{62}$ the fourth- and sixth-order cumulants of the net-proton distributions measured in experiments \cite{Adam:2020unf,STAR:2021rls}, which serve as the proxies for the net-baryon number fluctuations, have been presented for comparison. It is found that the theoretical results of $R^{B}_{42}$ are consistent with the measured $R^{p}_{42}$ within errors, except the data point at the collision energy  $\sqrt{s_{\mathrm{NN}}}$=14.5 GeV. In particular, a non-monotonic dependence of $R^{B}_{42}$ on the collision energy is observed in the low energy regime, which is in favor of experimental measurements. Note that the theoretical error indicated by the green bands is highly correlated, and the error bands are in fact comprised of a family of lines with a similar behavior to the central line, cf. \cite{Fu:2021oaw} for more details. The non-monotonic behavior seemingly also happens in the cases of other fluctuations, e.g., $R^{B}_{62}$, $R^{B}_{51}$ and $R^{B}_{82}$. A definite conclusion, however, has not yet been arrived at, due to the large errors in the regime of low collision energy for these fluctuations. The non-monotonic behavior of fluctuations on $\sqrt{s_{\mathrm{NN}}}$ arises from two facts: One is what we have mentioned above, viz., the chiral crossover becomes sharper and the oscillating amplitude of baryon number fluctuations is enhanced significantly with the increasing $\mu_B$. The other is that the freeze-out curve, as shown in the left panel of \Fig{fig:phasediagram}, deviates away from the crossover line and drops a bit at large $\mu_B$. In the plot of $R^{B}_{62}$ and $R^{p}_{62}$, one finds that the theory and experiment are compatible with each other within the relatively large errors. Interestingly, we find negative values for $R^{B}_{62}$ with the three collision energies. Moreover, our predictions of baryon number fluctuations of odd-orders as well as the eighth-order in \Fig{fig:Rm2Rm1-sqrtS}, await experimental confirmation in the near future.

\section{Summary}
\label{sec:summary}

In this work we have computed high-order baryon number fluctuations at finite temperature and density in a QCD-assisted LEFT within the fRG approach. Our results for fourth- and sixth-order baryon number fluctuations meet the lattice benchmarks at $\mu_B=0$, and are consistent with experimental measurements, and in particular, the non-monotonic dependence of the kurtosis of net-baryon number distributions on the collision energy is observed in our calculations. This non-monotonicity arises from the increasingly sharpened chiral crossover with the decrease of collision energy. The LEFT results are currently extended to computations in functional QCD, which will allow us to access the whole STAR, CBM and HADES energy range.

\bibliographystyle{JHEP}
\bibliography{ref-lib}

\end{document}